\documentclass{article}
\usepackage[utf8]{inputenc}
\usepackage{graphicx}
\usepackage{rotating}
\usepackage{subfig}
\usepackage{amsmath}
\title{Optimal Switching of Controlled Rectifiers}
\author{Shravan Mohan \\\small \textit{Santa Fe Research, Chennai}}
\date{February 2020}

\begin{document}

\maketitle

\begin{abstract}
    \noindent This paper discusses a linear programming approach for designing switching signals for controlled rectifiers to achieve a low input current \& output voltage total harmonic distortions. The focus here is on fully controlled rectifiers made with four-quadrant MOSFET based switches. This topology, unlike thyristor based rectifiers, can be turned ON or OFF anytime.  Yet another assumption made here is that the current drawn by the load is constant. The basic idea for designing the waveform is to first  time-discretize its one period. This discretization, along with Parseval's identity lead to a linear programming formulation for minimizing a weighted sum of total harmonic distortions of the input current and the output voltages. The LPs so obtained can be solved efficiently using standard solvers to obtain the switching instants. The method can be used for both single phase and three-phase rectifiers. Simulations are provided for corroboration.
\end{abstract}

\section{Introduction}
A rectifier is a power electronic device which converts AC power to DC power. Rectifiers are used in several low and high power applications; examples range from  consumer electronics to electric locomotives. The basic working principle of a rectifier is to invert parts/sections of the input AC waveform such that the output has a desired constant polarity. This inversion is achieved  through semiconductor based switches. The operation of the rectifier can be understood with the help of Figure 1. Note that if switches 1 and 4 are ON (switches 2 and 3 are OFF) during the first half of the input sinusoid, the output voltage across the load will be positive. Whereas, if switches 2 and 3 are ON (switches 1 and 4 are OFF), the output voltage across the load is inverted such that its polarity remains constant. Note that one could have switched on the respective pairs of switches only for certain parts of the period as compared to the aforementioned case. This way certain characteristics of the output voltage could have been controlled. On the other hand, in diode based full bridge rectifiers, forward biased diodes automatically turn ON/OFF to maintain constant polarity of output voltage, thereby providing no control. The semi-conductor switches used in controlled rectifiers are typically comprise of Thyristors/MOSFETs. Thyristor based rectifiers provide for turning ON capability (by sending a pulse to the gate) anytime they are forward biased, while only turn OFF once the current dies down to 0. This implies that the control is partly constricted. However, configurations such as the MOSFET plus diode based switch shown in Figure 2 allow for four-quadrant control; meaning that any switch can be switched ON or OFF anytime as long as the current drawn by the load is unidirectional and non-zero. The focus of this paper will be devising switching schemes for these fully controllable full-bridge rectifiers. In addition to controlling the output voltage, one might also want to design switching for minimizing the total harmonic distortion of the output voltage. Total harmonic distortion (THD), conventionally, is defined as the ratio of the total energy in the undesired harmonics to the total energy in the signal. Minimizing THD of the output voltage implies that the voltage fluctuation post the low pass filter. Note that voltage ripple at the output of the low pass filter is an important performance consideration in many practical applications. In addition to this, note that the input current drawn by the rectifier has a switched pattern as a result of  the switching of the rectifier switches. The switched current waveforms implies that the current drawn from the grid would be comprised of higher order harmonics, which is undesirable from a power system stability perspective. This effect might be more pronounced when several electrical vehicles might recharge off the grid, drawing several in range of several amperes per vehicle. Therefore, the goal, in addition to minimizing THD of the output voltage, must include minimizing the THD of the input current as well. Since there are two objectives, a natural choice would be to minimize a weighted sum of THD of output voltage and that of the input current. This would be the primary objective of the algorithm proposed in this paper. Note that the aforementioned discussion was with respect to single phase rectifiers. In industrial applications, three phase rectifiers are used since it can provide for more power for the same current ratings. It will shown later that the algorithm designed for single phase can be altered, appropriately, to cater to three phase AC inputs. \\\\
\begin{figure}[t]
    \centering
    \vspace*{-2cm}
    \hspace*{0.75cm}\includegraphics[scale=0.7]{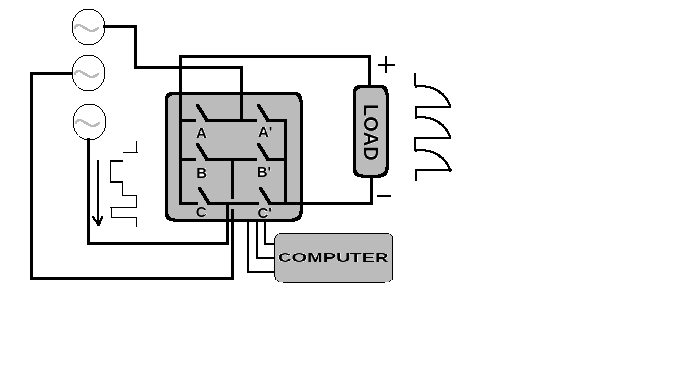}
    \vspace*{-2cm}
    \caption{A schematic representing a typical rectifier system. It shows a full bridge 3-phase rectifier which is assumed to be fully controllable, as discussed in the beginning of this paper. The switching of the gates is controlled by an external computer. The figure also depicts the load  voltage waveform. As it is assumed that the current drawn by the load is constant (or almost constant in the presence of a large inductance), the current drawn from the grid turns out to be a switched waveform as shown in this figure. }
    \label{fig:my_label}
\end{figure}
\noindent Conventionally, the problem of THD in currents drawn from the grid is mitigated using Active Front End (AFE) converters. The AFE uses an inductor on the grid side, which essentially suppresses the dynamics at high frequencies. In addition to this, the IGBT based rectifier legs (in single phase) are switched with sine-triangle PWM methodology. In fact, the PWM are given out of phase to the two legs so that the harmonic at the switching frequency is eliminated. Moreover, the  frequency of the triangular wave is also set to be an odd multiple of the sinusoid, thereby eliminating the odd harmonics as well. This method, although simple to understand and implement, has its disadvantages, and its performance can be improved using the proposed method in this paper. Firstly, the inductor on the grid side damps the transient response of the converter to varying loads. The inductor size can possibly be reduced by designing an switching scheme other than the usual sine-triangle PWM. Secondly, applying phase shifted versions of sine-triangle PWM only provides for one control parameter, with which only the switching frequency can be eliminated. In case more harmonic need to be eliminated, parallel converters, supplied by a dual  output transformer, are used. This increases the hardware costs and reduction in efficiency due to enhanced switching losses. In contrast to the aforementioned methods, the one presented in this paper is utilizes the degrees of freedom provided by the choices of switching angles for the IGBT legs to reduce the THD, even while maintaining the required output voltage. In addition to this, one can also perform harmonic elimination on both the output voltage and/or the input current. As the method is based on a linear program, the formulation is easy to grasp, mathematically tractable (determines if there is a solution with the given requirements or not) and is also amenable to implementation on a real-time system. \\\\
The following technical assumptions are made in this paper:
\begin{itemize}
    \item The switchings can happen only at specified time instants, which are equally spaced apart. This assumption arises from the fact that switching signal can be generated only at uniformly sampled time instants, as most computational devices work that way. 
    \item The switchings are assumed to be instantaneous, meaning that there is no delay between the change in switching signal and the output state of the switch. The time discretization. 
    \item The algorithm will not consider the dead-time (time gap between switching over to a different leg) in the analysis. The dead-times are typically small and here, the effect of dead-times on the output can be estimated. 
    \item The voltage drop across the switch while its in ON state will not be considered in the design of the switching scheme, since this is very small as compared to the AC/DC voltage levels.
\end{itemize}
\begin{figure}[t]
    \centering
    \hspace*{0.75cm}\includegraphics[scale=0.7]{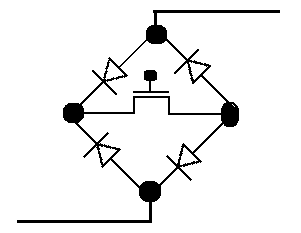}
    \caption{A common topology of a fully controllable switch. It can be seen that as long as the MOSFET is in OFF state, no current can flow through. If the MOSFET is in the ON state, then the current can flow in either directions (through the appropriate diodes). Note that a thyristor, on the other hand, can only be turned ON, while it only turns OFF once the current passing through it becomes zero.}
    \label{fig:my_label}
\end{figure}
\section{The Algorithm}
This section will discuss the construction of the LPs which will form the main algorithms for generating optimal switching schemes for single and three phase rectifiers. The constraints and the cost function of the LPs will be constructed and qualitative aspects of the solutions will be discussed.
\subsection{Single Phase Rectifier}
The algorithm shown in Figure \ref{fig:mainalgo_single} is the main algorithm determining the optimal switching scheme for a fully controlled  single phase rectifier. The cost function has two terms: the first term measures the total energy of the input current signal and the second terms measures the total energy in the output voltage signal. These terms are similar to the total energy term used in \cite{mohan2020linear}. The choice of the weight $\lambda$ on the user/application depending on the relative importance between the THDs of the input current versus that of the output voltage.\\\\
\noindent For the constraints part, the first equation defines the switching signal $x$ as a multiple of the a stochastic matrix $Z$ and a vector $S=[-1,~0,~1]$. This construction is again similar to the one used in \cite{mohan2020linear}. Note that the zero element corresponds to the free-wheeling operation of the rectifier. The second set of equations enforce the stochastic constraints on $Z$. The third constraint ensures that certain higher order harmonics of the input current waveform are all zero. The fourth constraint implies that the DC value of the output voltage is equal to the desired value. The fifth set of constraints bind the value of certain higher order harmonics of the output voltage to desired values (possibly zero). The details of the quality of the solution of the linear program and the effect of quantization can be found in \cite{mohan2020linear}.
\subsection{Three Phase Rectifier}
The algorithm shown in Figure \ref{fig:mainalgo_three} is the main algorithm determining the optimal switching scheme for fully controlled  three phase rectifier. As in the single phase case, the cost function has two parts to it: the first term is the total energy of the input current waveforms (through the three phases) and the second term is the total energy of the output voltage waveform. The term $\lambda$ is user defined and is a trade-off between the input currents THD (through the three phases) and the output voltage ripple. Note that the second term of the cost function uses appropriate coefficients derived from the three line voltages.\\\\
The construction of the constraints is a bit more intricate as compared to the single phase case. The vectors  $Z_{12}\times S$, $Z_{23}\times S$ and $Z_{31}\times S$ represent the choice of the state of the each of the three legs of the rectifier. If $Z_{12}\times S=1$ at any time index, this implies that the instantaneous current flows through the first phase, then the load and finally out of the second phase. Similar interpretations hold for the other cases. Also note that, as in the single phase case, the element 0 in the vector $S$ signifies free-wheeling operation mode of the rectifier. Hence, the first set of constraints. Note that one of the elements of the vectors of a particular row index can be non-zero.  Else, this would lead to a short circuit between different phase voltages. Hence the third set of constraints. Now, the second set of constraints enforce the that the average current drawn from either of the phases is equal to zero. The fourth set of constraints eliminate a bunch of harmonics from the input current. The fifth set of constraints ensure that the DC value of the output voltage is equal to its desired value. Finally, the last set of constraints bind the certain harmonics of the output voltage to desired values (possibly zero).   
\begin{figure}[t]
\centering
\fbox{\begin{minipage}{36em}
\begin{equation}
\begin{array}{l}
{\displaystyle \mbox{Solve}~~\min_{Z\in \mathcal{R}^{N\times m} }} ~~~\displaystyle\frac{1}{N}\left(\textbf{1}^{1\times N}\times Z\times S_p^\top \right) + \lambda \frac{1}{N}\left(S^{s,1}_p \times Z\times S_p^\top \right) \\\\
\displaystyle \mbox{sub to}~\left\{ \begin{array}{l}
 x = Z\times S,\\\displaystyle 
\displaystyle   Z \geq 0, ~~Z\times \textbf{1}^{m\times 1} = \textbf{1}^{N\times 1} \\ \displaystyle 
   f_k^\top x = 0, ~ \forall k\in \{k_1,\cdot,k_r\},\\ \displaystyle 
  \frac{1}{N} \left(S^{s,1}\right)^\top  x=\rm{DC}, \\ \displaystyle 
   f_l^\top \mbox{diag}\left(S^{s,1}\right) x=g_l^c+jg_l^s, ~ \forall l\in \{l_1,\cdot,l_z\}.
         \end{array} \right.
\end{array}
\label{eqn:MainLP_single}
\end{equation}
~~where $S=\left[-1,~0,~1\right]^\top $, $S_p=\left[1,~0,~1\right]$ and  $S^{s,1}=\left[\cdots~,~\sin\left(\frac{2\pi k}{N}\right)~,~\cdots \right]$, $S^{s,1}_p=\left[\cdots~,~\sin^2\left(\frac{2\pi k}{N}\right)~,~\cdots \right], 0\leq k\leq (N-1).$\\\\
\hspace*{1.5cm} {Clamp elements of $x$ to the nearest value in $S$}.
\end{minipage}}
\caption{The main optimization problem for determining the switching scheme for single phase full bridge rectifier.}
\label{fig:mainalgo_single}
\end{figure}

\begin{figure}[t]\hspace*{-1.5cm}
\fbox{\begin{minipage}{43em}
\begin{equation}
\begin{array}{l}
{\displaystyle \mbox{Solve}~~\min_{Z\in \mathcal{R}^{N\times m} }} ~~~\displaystyle  \frac{\textbf{1}^{1\times N}}{N}\left(Z_{12}+Z_{23}+Z_{31}\right)\times S_p^\top +~... \\ \hspace*{3cm}\displaystyle ...~\frac{\lambda}{2} \left(S^{s,12}_p\times Z_{12} + S^{s,23}_p\times Z_{23} + S^{s,31}_p\times Z_{31}\right)\times\textbf{1}^{N\times 1}\\\\
\displaystyle \mbox{sub to}~\left\{ \begin{array}{l}
 x_1 = \left(Z_{12} + Z_{31}\right)\times S,~ x_2 = \left(Z_{12} + Z_{23}\right)\times S ~\&~ x_3 = \left(Z_{23} + Z_{31}\right)\times S,\\
 \textbf{1}^{1\times N} x_1 = 0,~\textbf{1}^{1\times N} x_2 = 0,~ \textbf{1}^{1\times N} x_3 = 0,
 \\
 \displaystyle  Z_{12} \geq 0,~Z_{23} \geq 0,~Z_{31} \geq 0, ~~\left(Z_{12}+Z_{23}+Z_{31}\right)\times \textbf{1}^{m\times 1} = \textbf{1}^{N\times 1}, \\ \displaystyle 
   f_k^\top x = 0, ~ \forall k\in \{k_1,\cdot,k_r\},\\ \displaystyle 
  \frac{1}{N}\left( \left(S^{s,12}\right)^\top Z_{12} + \left(S^{s,23}\right)^\top Z_{23} + \left(S^{s,31}\right)^\top Z_{31} \right)\times s=\rm{DC}, \\ \displaystyle  f_l^\top \left( \left(S^{s,12}\right)^\top Z_{12} + \left(S^{s,23}\right)^\top Z_{23} + \left(S^{s,31}\right)^\top Z_{31} \right)\times s=g_l^c+jg_l^s, \\~ \hspace*{8.5cm}\forall l\in \{l_1,\cdot,l_z\}.
         \end{array} \right.
\end{array}
\label{eqn:MainLP_three}
\end{equation}
~~where $S=\left[-1,~0,~1\right]^\top $, $S_p=\left[1,~0,~1\right],$ and  $S^{s,12}=\left[\cdots~,~\sin\left(\frac{2\pi k}{N}\right) - \sin\left(2\pi/3+\frac{2\pi k}{N}\right)~,~\cdots \right]$, $S^{s,12}_p=\left[\cdots~,~\left(\sin\left(\frac{2\pi k}{N}\right) - \sin\left(2\pi/3+\frac{2\pi k}{N}\right)\right)^2~,~\cdots \right],$  and  $S^{s,23}=\left[\cdots~,~\sin\left(2\pi/3+\frac{2\pi k}{N}\right) - \sin\left(4\pi/3+\frac{2\pi k}{N}\right)~,~\cdots \right]$, $S^{s,23}_p=\left[\cdots~,~\left(\sin\left(2\pi/3+\frac{2\pi k}{N}\right) - \sin\left(4\pi/3+\frac{2\pi k}{N}\right)\right)^2~,~\cdots \right],$  and  $S^{s,31}=\left[\cdots~,~\sin\left(4\pi/3+\frac{2\pi k}{N}\right) - \sin\left(\frac{2\pi k}{N}\right)~,~\cdots \right]$, $S^{s,31}_p=\left[\cdots~,~\left(\sin\left(4\pi/3+\frac{2\pi k}{N}\right) - \sin\left(\frac{2\pi k}{N}\right)\right)^2~,~\cdots \right], 0\leq k\leq (N-1).$\\\\
\hspace*{1.5cm} {Clamp elements of $x$ to the nearest value in $S$}.
\end{minipage}}
\caption{The main optimization problem for determining the switching scheme for three phase full bridge rectifier.}
\label{fig:mainalgo_three}
\end{figure}

\noindent \newline \textbf{\underline{Remark on Non-constant Current Loads}:} \textit{There is a possibility that the loads might not have a large inductive component, implying that the variability in the current drawn might be large. In that case, the assumption of (almost) constant current drawn at the output is invalid and this would lead to a sub-optimal switching scheme.  The THD calculation of the input current in that case is not straightforward, as one has to account for the attenuation in frequency components brought about by the load. This is akin to minimizing a weighted total harmonic distortion of the input current, of course weighted appropriately as per the load. However, minimizing weighted THD is a challenging problem and is not within the scope of this paper. In case the current waveform is measured and varies much slowly as compared to the fundamental time period of the AC input, then that might be taken into account for THD calculations in the cost function.}\\\\
\textbf{\underline{Remark on Transient Performance}:} \textit{Note that its the steady state premise under which the algorithm is built upon. The algorithm does not prescribe a transition methodology from one steady state to the other. The transient response must not only be fast but possibly, also not lead to overshoots. Needless to say, designing the transition methodology would need the knowledge of the load dynamics. One simple trick would be to replace a section of the waveform with what is to be, and then increase this section step by step so as the achieve the desired transient response. However, a more detailed approach is not within the scope of this paper.}  
\\\\
\textbf{\underline{Remark on Unbalanced Input Voltage Supplies}:} \textit{In practice, it is conceivable that the inputs voltages are not balanced across the three phases, and possibly are contaminated with higher order harmonics. However, the method prescribed in this paper subsumes these possibilities. All that one needs to do is to change the cost function and the constraints appropriately to reflect the actual input voltage waveform, instead of the unit amplitude sinusoid used in the formulation.}  
\section{Simulations}
This section presents simulation results for  both single phase and three phase rectifiers, under different output voltage and input current harmonic requirements. Each simulation result has four components: (i) the switching pattern for each leg, (ii) the output voltage waveform, (iii) the magnitude Fourier spectrum and (iv) the low-pass filtered version of the output voltage. The switching pattern can be seen to have three levels: the switches S1 and S4 are ON when the level is equal to 1, the switched S2 and S3 are ON when the level is equal to -1 and all the switches are OFF while the level is equal to zero. When all the switches are OFF, the load current flows through the free-wheeling diodes under (almost) potential difference across the diodes. In this condition, the input current drawn is zero and the voltage across the load is assumed to be zero. Note that the switching scheme is essentially the input current waveform, except for a scaling factor. The same logic applies to three phase rectifiers as well. The switching schemes for the three sets of legs (corresponding to each phase) can be deciphered from the different colour plots. The output voltage can be seen to be in line with the intuition that it must be comprised of chunks of the sinusoid, or its inversion. The Fourier components are calculated using FFT and it can be seen that the magnitude reduces as one progresses along the harmonics, as is the case with most switched signals. The low-pass filter is chosen to have a cut-off frequency at 50Hz, and the output of the filter can be seen to hover around the desired DC voltage. It must be noted that the higher order harmonics and the voltage ripple are subdued in the three phase rectifiers, as compared to single phase rectifiers. The higher order harmonics are subdued because intuitively, three legs provide for more degrees of freedom and hence the THD can be reduced further. The voltage ripple post low-pass filtering is subdued because the higher order harmonics start from a larger value as compared to the single phase rectifier case. This implies that the same low-pass filter has a stronger attenuation on these harmonics. In all the simulation cases, the switching schemes are designed to eliminate the second, the fourth and the sixth harmonics from the output voltage. Finally, note that there is a possibility that there may not exist switching schemes which adhere to all harmonic requirements on the input current and the output voltage. But infeasible cases can be easily determined as the optimization problem is a linear program. All the optimizations are performed using CVXPY on Python \cite{diamond2016cvxpy}.
\begin{sidewaysfigure}
  \centering
  \subfloat{\includegraphics[scale=0.575]{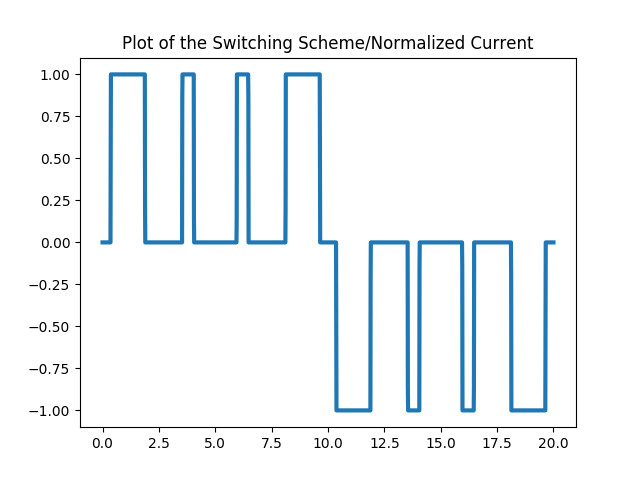}} \quad
  \subfloat{\includegraphics[scale=0.575]{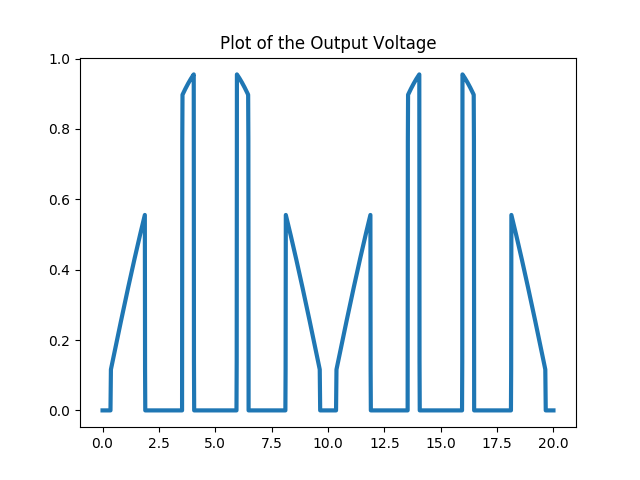}}\quad
  \subfloat{\includegraphics[scale=0.575]{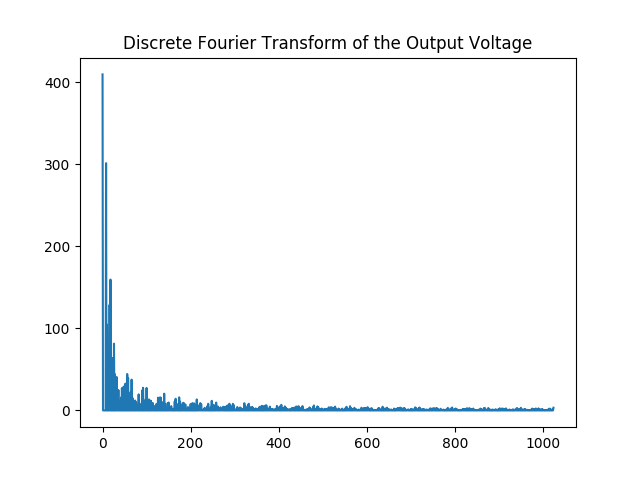}}\quad
  \subfloat{\includegraphics[scale=0.575]{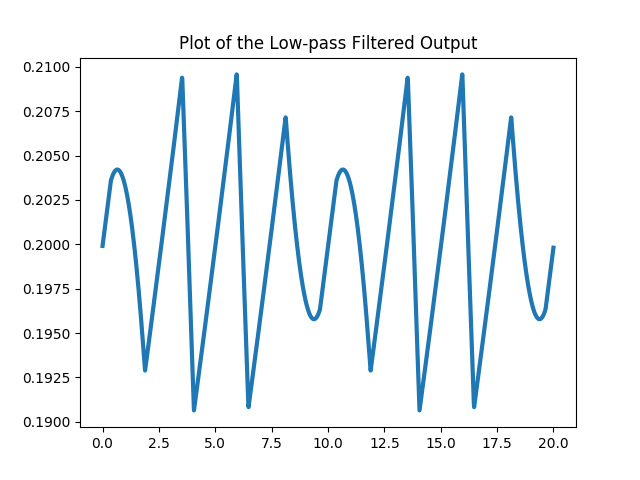}}
  \caption{This figure presents simulation results for the case with the input sinusoidal amplitude equal to 1, the desired output DC voltage set to 0.2, the second, the fourth and the sixth harmonics of the output voltage eliminated, , and the $\lambda$ set to 10.  The top left subfigure shows the current waveform, which is switched as expected. The top right subfigure is the output voltage waveform, and the bottom left subfigure is its magnitude spectrum. The bottom right figure shows the filtered output voltage, with a RL (R=1$\sigma$ and L=20 mH) low-pass filter.
  \label{fig:single_phase_1}}
\end{sidewaysfigure}

\begin{sidewaysfigure}
  \centering
  \subfloat{\includegraphics[scale=0.575]{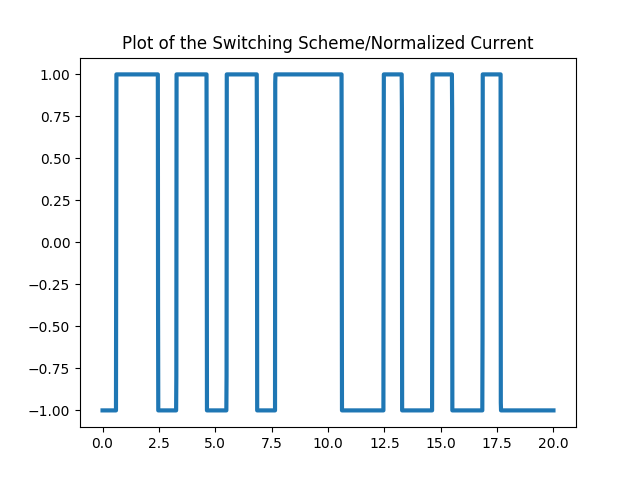}} \quad
  \subfloat{\includegraphics[scale=0.575]{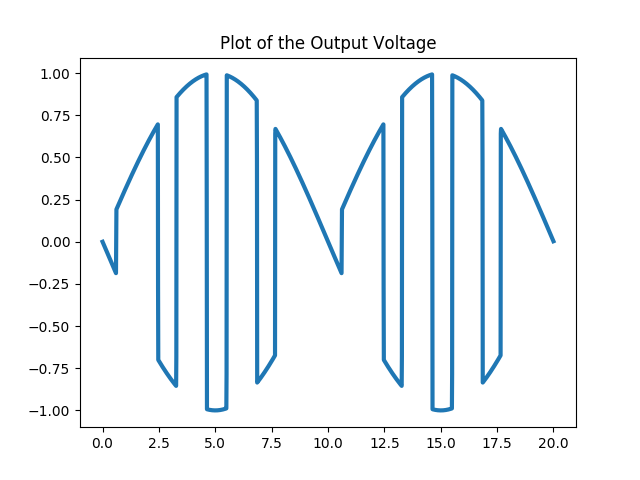}}\quad
  \subfloat{\includegraphics[scale=0.575]{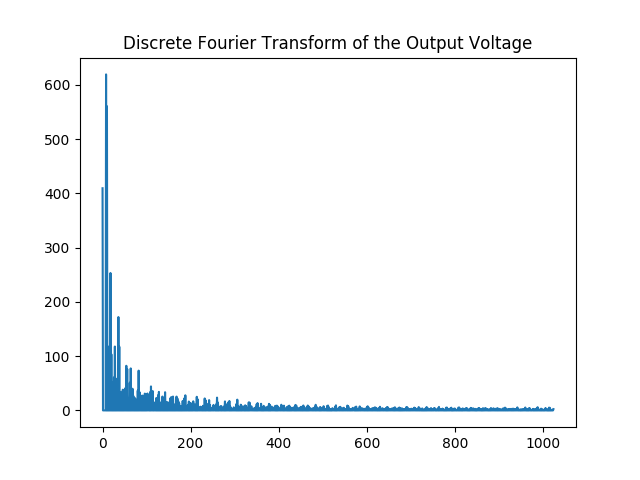}}\quad
  \subfloat{\includegraphics[scale=0.575]{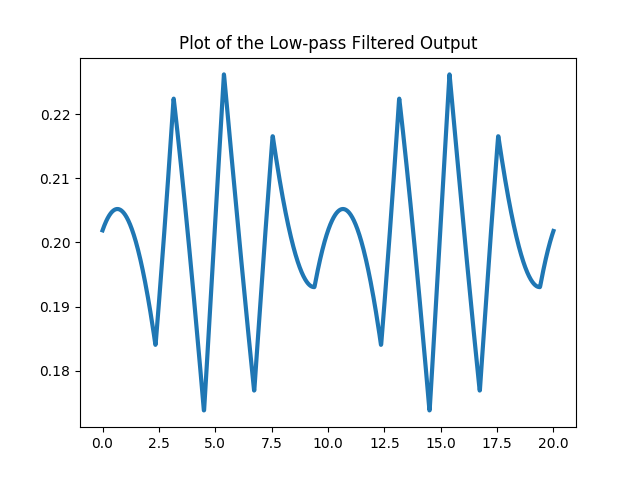}}
  \caption{This figure presents simulation results (without allowing for free-wheeling operation) for the case with the input sinusoidal amplitude equal to 1, the desired output DC voltage set to 0.2, the second, the fourth and the sixth harmonics of the output voltage eliminated, and the $\lambda$ set to 10. The top left subfigure shows the current waveform, which is switched as expected. The top right subfigure is the output voltage waveform, and the bottom left subfigure is its magnitude spectrum. The bottom right figure shows the filtered output voltage, with a RL (R=1$\sigma$ and L=20 mH) low-pass filter.
  \label{fig:single_phase_2}}
\end{sidewaysfigure}

\begin{sidewaysfigure}
  \centering
  \subfloat{\includegraphics[scale=0.575]{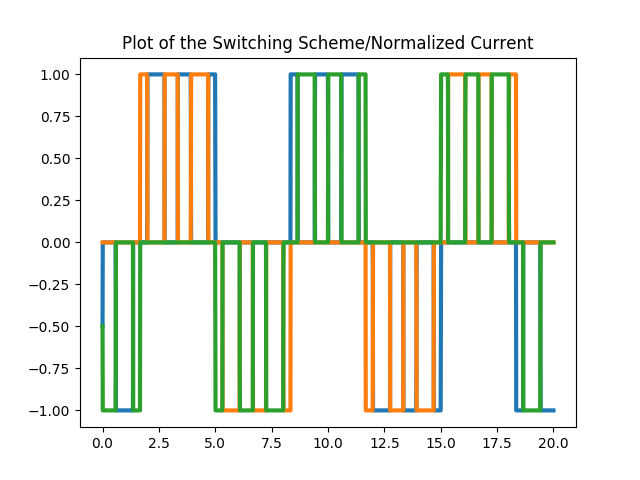}} \quad
  \subfloat{\includegraphics[scale=0.575]{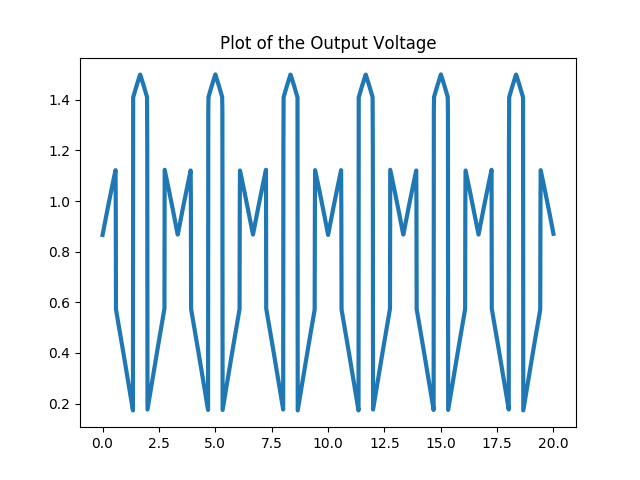}}\quad
  \subfloat{\includegraphics[scale=0.575]{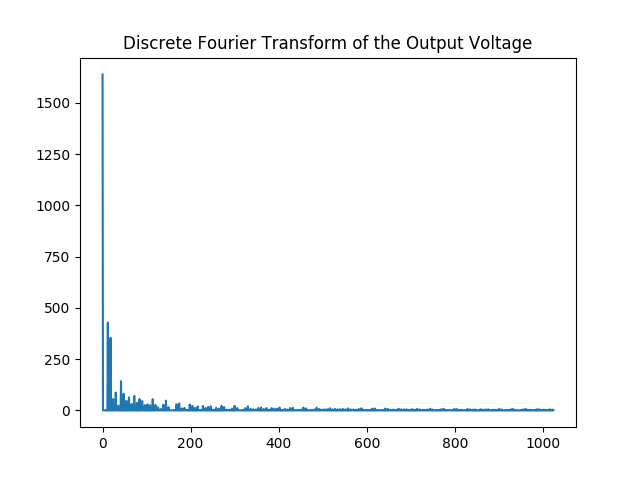}}\quad
  \subfloat{\includegraphics[scale=0.575]{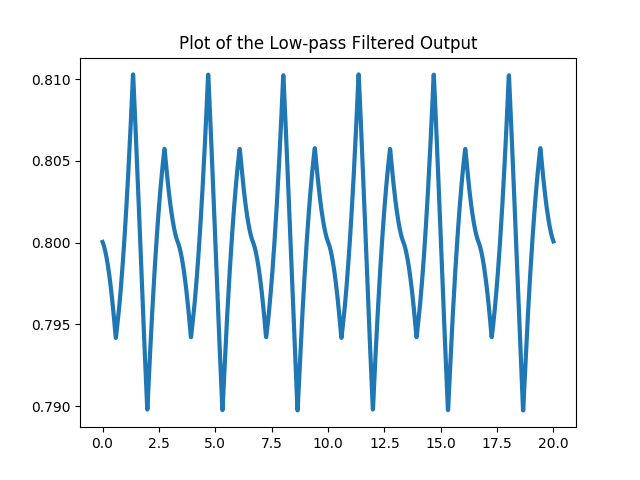}}
  \caption{This figure presents simulation results for the case with the three phase input sinusoidal amplitude equal to 1 (without the free wheeling operation), the desired output DC voltage set to 0.8, the second, the fourth and the sixth harmonics of the output voltage eliminated, and the $\lambda$ set to 10. The top left subfigure shows the current waveform, which is switched as expected. The top right subfigure is the output voltage waveform, and the bottom left subfigure is its magnitude spectrum. The bottom right figure shows the filtered output voltage, with a RL (R=1$\sigma$ and L=20 mH) low-pass filter.}
  \label{fig:three_phase_1}
\end{sidewaysfigure}

\begin{sidewaysfigure}
  \centering
  \subfloat{\includegraphics[scale=0.575]{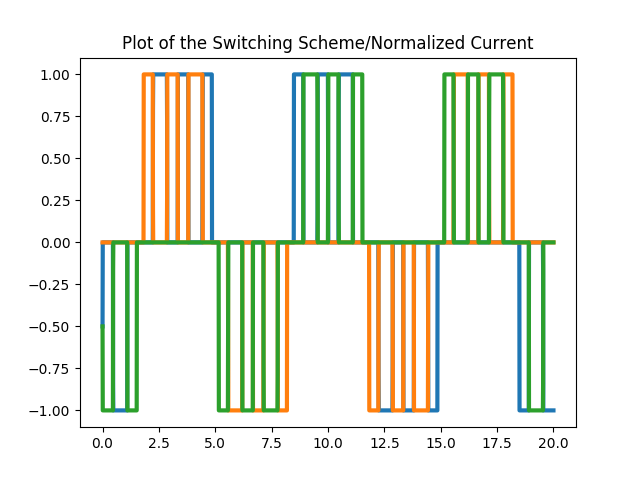}} \quad
  \subfloat{\includegraphics[scale=0.575]{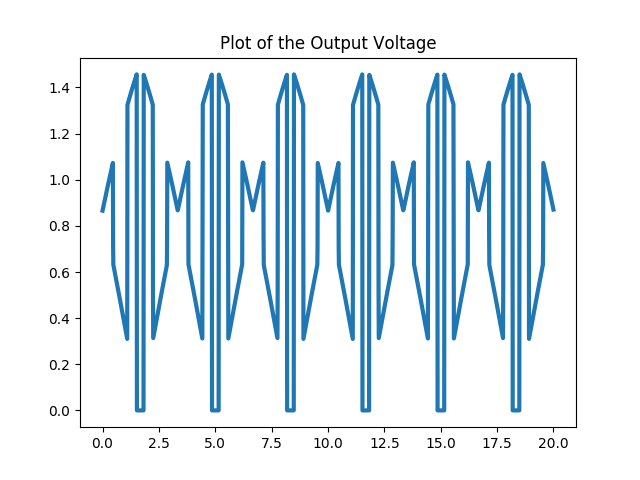}}\quad
  \subfloat{\includegraphics[scale=0.575]{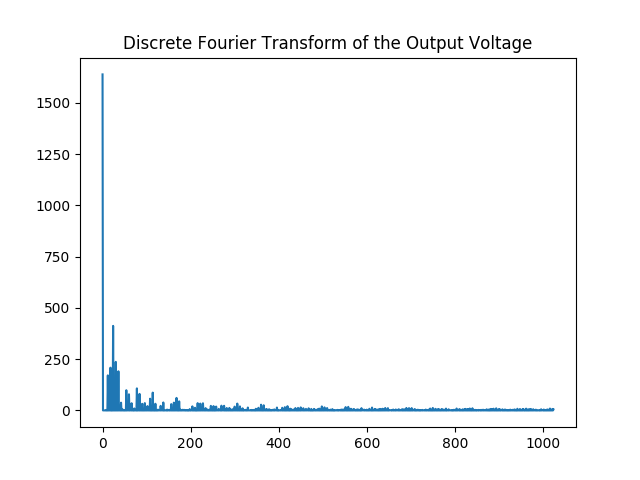}}\quad
  \subfloat{\includegraphics[scale=0.575]{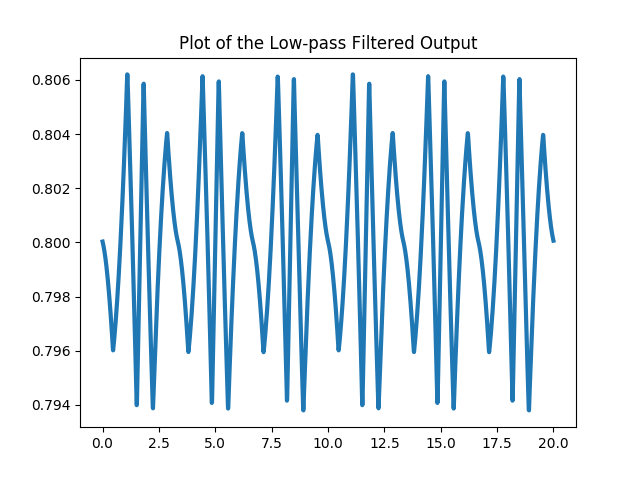}}
  \caption{This figure presents simulation results for the case with the three phase input sinusoidal amplitude equal to 1, the desired output DC voltage set to 0.8, the second, the fourth and the sixth harmonics of the output voltage eliminated, and the $\lambda$ set to 10. The regularization factor in the cost function is set to 10. The top left subfigure shows the current waveform, which is switched as expected. The top right subfigure is the output voltage waveform, and the bottom left subfigure is its magnitude spectrum. The bottom right figure shows the filtered output voltage, with a RL (R=1$\sigma$ and L=20 mH) low-pass filter.}
  \label{fig:three_phase_2}
\end{sidewaysfigure}

\section{Conclusions}
This paper proposes a linear programming approach for designing switching schemes for fully controllable rectifiers. The objective is to minimize the THD of the output voltage and that of the input current drawn from the grid. The objective can also be enhanced with requirements of harmonic elimination the output voltage and/or input current. The algorithm proposed is easily extended to the three phase rectifier case. The proposed method essentially uses the degrees of freedom provided by the possible choices of switching schemes to achieve an optimal performance. The use of LP implies that the optimization can be solved easily on a computer and is also amenable for real-time implementations. Simulations for various cases were also presented, which corroborate the proposed idea.  

\bibliographystyle{alpha}
\bibliography{refs}

\end{document}